\documentclass[12pt]{article}
\usepackage{amsmath,amssymb,graphicx,mathrsfs,hyperref}

\newcommand{\be}{\begin{equation}}
\newcommand{\ee}{\end{equation}}
\newcommand{\bea}{\begin{eqnarray}}
\newcommand{\eea}{\end{eqnarray}}

\def\({\left(} \def\){\right)}

\begin{document}
\title{\vspace{-1.8in} Proof of a Universal Lower Bound on the Shear Viscosity to Entropy Density Ratio}
\author{\large Ram Brustein ${}^{(1)}$,  A.J.M. Medved ${}^{(2)}$  \\ \ \\
(1) Department of Physics, Ben-Gurion University, \\
    Beer-Sheva 84105, Israel   \\
(2)    Physics Department,  University of  Seoul \\
  Seoul 130-743  Korea \\
    E-mail: ramyb@bgu.ac.il,\ allan@physics.uos.ac.kr }
\date{}
\maketitle

\begin{abstract}

It has been conjectured, on the basis of the gauge-gravity duality, that the ratio of the shear viscosity to the entropy density  should be universally bounded from below by $1/4\pi$ in units of the Planck constant divided by the Boltzmann constant. Here, we prove the bound for any ghost-free extension of Einstein gravity and the field-theory dual thereof. Our proof is based on the fact that, for such an extension, any gravitational coupling can only increase from its Einstein value. Therefore, since the shear viscosity is a particular gravitational coupling, it is minimal for Einstein gravity. Meanwhile, we show that the entropy density can always be calibrated to its Einstein value. Our general principles are demonstrated for a pair of specific models, one with ghosts and one without.


\end{abstract}
\newpage

Kovtun, Son and Starinets (KSS) have proposed that the ratio of  the shear viscosity $\eta$ to the entropy density $s$ should be universally bounded from below by $1/4\pi$ in units of the Planck constant divided by the Boltzmann constant $\hbar/k_B$ \cite{KSS-hep-th/0405231}.
In kinetic theory, the shear viscosity of a fluid is directly proportional to the mean free path of the quasiparticles, suggesting that $\eta/s$ is much larger than $\hbar/k_B$ in weakly coupled fluids for which the mean free path of the quasiparticles is always large. Thus, a bound on $\eta/s$ is mostly relevant to strongly interacting fluids. The uncertainty principle can be used in this context to argue that the ratio $\eta/s$ in units of    $\hbar/k_B$ is bounded from below by a constant of order unity
\cite{KSS-hep-th/0405231}.  So far, two classes of quantum fluids  are known to have values of $\eta/s$  that approach  $\frac{1}{4\pi}\frac{\hbar}{k_B}$: Strongly correlated ultracold Fermi gases and the quark--gluon plasma. (See \cite{Schafer:2009dj} for a recent review and many references.) Experiments in both systems \cite{coldFG,Collaboration:2009yd} have now reached the necessary precision to probe the KSS bound.

The gauge--gravity duality \cite{maldacena1,maldacena2} provides a hydrodynamic description of strongly coupled field theories in terms of the hydrodynamics of a black brane  in  an asymptotically anti-de Sitter (AdS) spacetime \cite{KSS-hep-th/0309213}.  It was shown in
\cite{PSS-hep-th/0104066} that, for strongly coupled field theories,
$\eta/s =1/4\pi$  when the bulk gravitational theory is Einstein's
(from here on, $\hbar,\;k_B,\;c=1$). This value has since been confirmed in many examples. (For references and further discussion,
see  \cite{SS-0704.0240}.)
However, recent findings have cast doubts over the universal nature of the bound. For instance, when the gravitational Lagrangian includes the square of the $4$-index Riemann tensor, the ratio $\eta/s$
can either be smaller or larger than its Einstein value
\cite{BLMSY-0712.0805,same-day}.  These modifications can be understood from the observation \cite{BM-0808.3498}  that $\eta/s$ is equivalently a ratio of  two different gravitational couplings, each associated with a differently polarized graviton \cite{BGH-0712.3206}. If the gravitational theory is Einstein's  or related to Einstein's by a  field redefinition, then the couplings will be independent of the polarization and
$\eta/s=1/4\pi$. In general, however, the couplings for differently polarized gravitons are distinct, and there is no longer any reason  to expect that  $\eta/s=1/4\pi$.

As will be explained, if we impose the physical requirement that extensions of Einstein gravity must be ghost free, then any gravitational coupling can only {\em increase} from its Einstein value. We will show, in particular, how this outcome applies to the shear viscosity. It will then be demonstrated how the entropy density for any extension can always be calibrated to its Einstein value. Combining both facts will allow us to establish that,  if the ratio $\eta/s$ does differ from the Einstein result, then it must necessarily be larger than $1/4\pi$.

We will consider a general theory of gravity in AdS whose action depends on
the metric $g_{\mu\nu}$, the Riemann tensor ${\cal R}_{\rho\mu\lambda\nu}$,
matter fields $\phi$ and their covariant derivatives:
$
I=\int\!\! d^{d+1} x \sqrt{-g}\
\mathscr{L}\left({\cal R}_{\rho\mu\lambda\nu},g_{\mu\nu},\nabla_\sigma
R_{\rho\mu\lambda\nu},\phi,\nabla\phi,\ldots\right)
$
with $d\geq 3$.
We will further assume the existence of stationary $(p+2)$-black brane solutions with a bifurcate Killing
horizon and described by the following metric:
$
ds^{2}=-g_{tt}(r)\,dt^{2}+g_{rr}(r)\,dr^{2}+g_{xx}(r)\,dx^{i}\,dx_{i}\;, \ i=1,\dots,p.
$
The black brane event horizon is at $r=r_h$, where $g_{tt}$ has a
first-order zero, $g_{rr}$ has a first-order pole and
all other metric components are finite. All the
metric components are taken to depend only on $r$ and, therefore, the
metric is Poincare invariant in the $(t,x_i)$ subspace. Also,
the AdS boundary is taken to be  at $r\to\infty$, where
the metric asymptotically approaches its AdS form.

Let us discuss small perturbations  of the brane metric:
$g_{\mu\nu}\rightarrow g_{\mu\nu} + h_{\mu\nu}$ with $h_{\mu\nu}\ll 1$.
We will take $z$ as  their propagating direction on the brane. It is well known  that, for a suitable choice of gauge \cite{PSS2},
the highest-helicity
polarization of the $h_{xy}$ gravitons decouples from all others. (Obviously, $x,y$ can be replaced by any other orthogonal-to-$z$ transverse dimensions.) A standard procedure \cite{Kovtun:2005ev,SS-0704.0240} that involves taking the hydrodynamic limit and using the Kubo formula  allows one to extract the shear viscosity  from the correlation function of the dissipative energy-momentum tensor $T_{xy}\sim \eta \partial_t h_{xy}$. As explained in \cite{BM-0808.3498}, this procedure is valid for extensions of Einstein gravity and amounts to extracting the gravitational coupling for the $h_{xy}$ gravitons. The very same procedure can also be implemented at any radial distance from the brane \cite{Iqbal:2008by,BM-shear}.  Applying it at the AdS boundary, one then learns about the shear viscosity in the dual field theory.

We can determine the $h_{xy}$ coupling by calculating the propagator in the one-particle exchange approximation, which is valid because $h_{xy}\ll 1$. For Einstein gravity, only massless spin-$2$ gravitons are exchanged, but gravitons can, for a general theory, be  either   massless or massive and of  either spin-$0$ or spin-$2$. Particles of any other spin, in particular vectors, can not couple linearly  to a  conserved source and so can safely be neglected when evaluating the propagator.  Accordingly, the graviton  propagator
$[{\cal D}(q^2)]^{\ \nu \; \beta}_{\mu \; \alpha}\equiv \langle h_{\mu}^{\ \nu}(q)
h_{\alpha}^{\ \beta}(-q)\rangle$
must be of the following irreducibly decomposed form
\cite{Zakharov,veltman,Dvali1,Dvali2}:
\bea
[{\cal D}(q^2)]^{\ \nu \; \beta}_{\mu \; \alpha} \;&=&\; \left(\rho_E(q^2) + \rho_{NE}(q^2)\right) \left[\delta^{\ \beta}_{\mu}\delta^{\ \nu}_{\alpha}- \frac{1}{2}\delta^{\ \nu}_{\mu}\delta^{\ \beta}_{\alpha}\right] \frac{G_E}{q^2}
 \nonumber \\
&+&\!\!\sum_i\rho^i_{NE}(q^2)\left(\delta^{\ \beta}_{\mu}\delta^{\ \nu}_{\alpha}- \frac{1}{3}\delta^{\ \nu}_{\mu}\delta^{\ \beta}_{\alpha}\right)\frac{G_E}{q^2-m_i^2}
\nonumber \\
&+&\!\!\sum_j {\widetilde\rho}^{\; j}_{NE}(q^2)\; \delta^{\ \nu}_{\mu}\delta^{\ \beta}_{\alpha}\;\frac{G_E}{q^2 - {\widetilde m}_j^2}\;.
\label{decomp}
\eea
Here, $G_E$ is Newton's constant and we have  denoted  the
Einstein  and ``Non-Einstein'' parts of the gravitational couplings $\rho$  by the subscripts $E$ and $NE$. We have separated the contribution of the massless spin-$2$ particles,  massive spin-$2$ particles with mass $m_i$ and scalar particles with mass ${\widetilde m}_j$. Some of the masses $m_i$, ${\widetilde m}_j$ may vanish or be parametrically small in certain cases.
The couplings or $\rho$'s  are dimensionless  quantities   and can depend  on the momentum scale $q$; in particular,  $\rho_E(0)=1$.

For a ghost-free theory, all of the couplings must remain positive at all energy scales \cite{Dvali1}; meaning that {\em the propagator can  only increase} relative to its Einstein value.

As discussed above, the shear viscosity for {\em any} theory of gravity can
be determined directly from the propagator
$\langle h_{xy}(q)h_{xy}(-q)\rangle $ when taken to the hydrodynamic limit.
 In this limit, the temperature $T$ is the largest relevant scale, so that both the energy $\omega$ and  momentum $\vec{q}\;$ have to satisfy
$\omega/T$, $|\vec{q}|/T\ll 1$ (with $\omega$ and $|\vec{q}|$  not necessarily of the same magnitude). For Einstein's theory, the above procedure yields the well-known answer $\eta_E =1/(16\pi G_E)$.  For a general theory,  the corrections can be read off the  propagator
in Eq.~(\ref{decomp}). As we only have to consider mode contributions  such that $m/T\to 0$,
the shear viscosity $\eta_X$ for a generic theory $X$  can be expressed
as follows:
\be
\frac{\eta_{X}}{\eta_E}\;=\;\left[\frac {\langle h_{x}^{\ y} h_{y}^{\ x}\rangle_{X}}
{\langle h_{x}^{\ y} h_{y}^{\ x}\rangle_E}\right]\;=\; 1+\frac{1}{\rho_E(0)}\sum\limits_i \rho_{NE}^{\; i}(0)\; .
\label{bonus}
\ee
The sum represents the non-Einstein  contribution of spin-2 particles to the $xy$-polarization channel
in Eq.~(\ref{decomp}). The scalars and the trace parts of the massless and massive gravitons do not contribute to the sum in Eq.~(\ref{bonus}).

Irrespective of the precise nature of the corrections,
we  know that $\rho_E(0)=1$ and that, for any ghost-free theory of gravity, the $\rho$'s must be positive. It follows that their effect can only be to increase $\eta$  relative to its Einstein value:
\be
\frac{\eta_{X}}{\eta_E}\ge 1\;.
\label{bonusI}
\ee
This lower bound must be true for any coordinate system or choice of field definitions, as the absence
of ghosts is an invariant statement that is insensitive  to these choices.
Further, as  previously discussed, this bound is valid at any radial distance from the brane and, in particular, near the AdS boundary at
$r\to\infty$. The gauge--gravity duality then implies that the shear viscosity of any field-theory dual also satisfies Eq.~(\ref{bonusI}). That is,  the smallest $\eta$ for a field theory must be for the theory dual to Einstein gravity.

To bring the discussion back to the ratio of interest $\eta/s$, let
us next consider the entropy density. The gauge--gravity duality tells us
that the entropy density for a given field theory is the same
as that of its black brane dual.
Also, the temperature $T$ of the
field theory can be identified with
the Hawking temperature of the black brane. The latter is fixed by the horizon radius $r_h$ (along with any charges)  in units of the AdS curvature scale and depends explicitly on the geometry but {\em not} on the underlying Lagrangian. Working at a fixed  value of temperature or fixed $r_h$,
we will demonstrate that, although the black  brane entropy $S_X$ for a generic gravity theory can be different from the Einstein value
$S_E$, it is always possible to find a field redefinition that calibrates the entropy {\em density} $s_X$ to the Einstein density $s_E$.

For a $(p+2)$-black brane with $p\geq 2$ transverse dimensions, the
entropy for Einstein's theory is
$
S_E\;=\; V_\bot {r_h^p}/{(4G_E)}\;,
\label{ein-ent}
$
where  $V_\bot$ is the transverse volume of the brane which includes  all other numerical factors and inverse factors of the AdS curvature radius.
For a general theory $X$ that extends Einstein gravity,  the entropy
is given by Wald's formula \cite{wald1,wald2,wald3} and can  be expressed as a correction to the Einstein value:
\be
S_X\;=\; \frac{V_{\bot} r_h^p}{4G_X} \;=\; \frac{V_{\bot} r_h^p}{4G_E+\lambda\delta G}\;+\;{\cal O}[\lambda^2]\;.
\label{gen-ent}
\ee
Here, $G_X$ is the gravitational coupling for $X$ and
we have fixed the horizon radius of the brane since our interest is to compare the theories at a fixed temperature. $\lambda$ is a parameter that controls the strength of the correction to Einstein's theory and $\delta G$ indicates the first-order (in $\lambda$) shift in the coupling.
One can determine $G_X$ and, hence, $\delta G$  by calculating the Wald  entropy  as prescribed in \cite{BGH-0712.3206} and then Taylor expanding
to any desired order of accuracy. Results are presented at the lowest
non-trivial order for simplicity and clarity.

Next, let us consider a conformal field redefinition of the  metric such that
$g_{\mu\nu}\rightarrow{\tilde g}_{\mu\nu}=e^{\Omega}g_{\mu\nu}\;$ with  $\Omega= - \frac{2}{p}\lambda\delta G/G_E$. This field redefinition should  be accompanied by an appropriate rescaling $G_X\to G_{\widetilde X}$
to preserve the form of the leading-order Lagrangian $(16\pi G)^{-1} \sqrt{-g}
{\cal R}$.
The position of the horizon  can  be determined as the largest root of the ratio $|g_{tt}/g_{rr}|$ and so will be left unchanged.
The entropy for the transformed theory is given by
\be
S_{\widetilde X}\;=\; \frac{{\widetilde V}_\bot\; r_h^p}{4G_{\widetilde X}} \;=\; \frac{\left(V_\bot-\lambda\frac{\delta G}{G_E}\right)r_h^p}{4G_E }\;+\;{\cal O}[\lambda^2]\;,
\label{gen-ent-2}
\ee
which, again, can be extended to any desired order in
$\lambda$.
The entropy is invariant under the field redefinition, as it must be.
It is also clear from Eq.~(\ref{gen-ent-2}) that
$S_{\widetilde X}/S_E={\widetilde V}_\bot/{V_\bot}$, and so
\begin{equation}
s_{\widetilde X}=s_E.
\label{ent-dens-gen}
\end{equation}

This last result and the gauge--gravity duality implies that the entropy densities of the dual field theories are also equal, while
Eq.~(\ref{bonusI}) tells us that their shear viscosities satisfy  $\eta_{\widetilde X}\geq \eta_{E}$.
Hence,
\be
\frac{\eta_X}{s_X} \;\geq \; \frac{\eta_E}{s_E} \;=\;\frac {1}{4\pi}\;,
\label{bound}
\ee
where the tildes have been dropped for what must be  a field-redefinition-invariant statement. We have thus {proved} the KSS bound for any consistent extension of Einstein gravity {and} its  field-theory dual.

Let us now discuss some examples. Obviously, in any theory which is equivalent to Einstein gravity with simple enough matter interactions, $\eta/s=1/4\pi$. This can occur for theories that contain only topological corrections such as  Gauss-Bonnet gravity in $4D$ and Lovelock gravity in higher even-numbered dimensions, or for  theories that can be brought into Einstein's by a field redefinition such as  $f({\cal R})$ gravity.

To obtain a gravity theory without ghosts that  extends Einstein gravity  in a non-trivial way,  one can start with a ghost-free theory and then integrate out some of the matter or  gravity degrees of freedom in a consistent way.  We have chosen to discuss a simple model in this class: Einstein's theory in $4+n$ dimensions, with the $n$
extra dimensions compactified on  a torus of radius $R$. From a
$4$-dimensional point of view, everything is trivial in the far infrared when the horizon radius is larger than the compactification scale or
$r_h\gg R$. But, when the energy scales are high enough to probe the compactified dimensions or $r_h\ll R$, the correct description is
the higher-derivative theory that is obtained after the Kaluza--Klein (KK) modes of the torus have consistently been integrated out.

From a $4D$ point of view, each extra dimension $i={1,2,\ldots n}$ induces an infinite tower of massive KK modes with uniformly spaced masses $m_{k_{i}}\sim k_i/R$ ($k_i=1,2,\ldots\infty$) for particles of spin-$0$, $1$ and $2$. Only the spin-$2$ particles will be relevant to the shear ($xy$) channel of  the  two-graviton propagator (\ref{decomp}), and this contribution  can  be expressed as the following  sum:
\be
 \left[{\cal D}_{KK}\right]^{\ y\ x}_{x\ y}\;\sim\;  R^2 G_E\sum_{i=1}^{n}\sum_{k_{i}=1}^{\infty}\frac{\rho_{k_{i}}(q^2)}
{R^2 q^2 - k_i^2}\; .
\label{kk-sum}
\ee
The coefficients $\rho_{k_i}$  must be non-negative for this ghost-free theory at  all energy scales.   Then,  using the translational symmetry of the torus, one finds
that ${\cal D}_{KK}$ is vanishingly small for  $q\ll 1/R$ (as anticipated), whereas
${\cal D}_{KK}\sim  R/q$ for $q\gtrsim 1/R$. In this latter case,
the KK contribution clearly dominates over the standard $1/q^2$ contributions. Recall that, for the hydrodynamic limit, the masses  contributing to  Eq.~(\ref{kk-sum}) need to be small
or $m\sim 1/R \ll T$, which  is indeed the case when $r_h\ll R$
since $r_h\sim 1/T$.

It is evident that,  for  scales $q\gtrsim 1/R$,
the shear viscosity  must increase  from its Einstein value.  On the other hand, the entropy density can always be computed
in either $4+n$ dimensions or just $4$, with the same result guaranteed.
This is true by virtue of the compactified dimensions making the same $R^n$
contribution to  both the area density in the numerator  and
the gravitational coupling  in the denominator.
Consequently, from a $4D$ point of view,  $\eta/s$ saturates the bound~(\ref{bound}) in the IR where the theory is Einstein's and then increases towards the UV due to the increase in $\eta$.

Our assertion is that the bound~(\ref{bound}) has to hold for a ghost-free theory. However, the converse is not true.
It is possible, as demonstrated below, that the bound holds for some  theories with ghosts but not for others; apparently, some ghosts are ``friendlier'' than others.
For concreteness, let us discuss  $5D$ Gauss--Bonnet gravity. The
Lagrangian density of this theory is
$
{\cal L} \;=\; {\cal R} +12 \;+ \;\lambda G_E\left[{\cal R}_{abcd}{\cal R}^{abcd}
-4{\cal R}_{ab}{\cal R}^{ab} + {\cal R}^2\right]
$
where we have set the AdS curvature radius equal to unity and $\lambda$ is a dimensionless constant. It is well known that this theory has ghosts for {\em any} value of $\lambda$, as one can  readily see by performing a field redefinition that retains  only the $4$-index (squared) correction $\sim {\cal R}_{abcd}{\cal R}^{abcd}$ and then computing the complete propagator.

The graviton propagator is calculated as follows:
We expand the metric $g_{\mu\nu}\rightarrow g_{\mu\nu}^{(0)}+h_{\mu\nu}$ (with a superscript of $(0)$ always denoting the $\lambda=0$ solution) and then calculate the graviton  kinetic terms which contain exactly two $h$'s and two derivatives. From the formalism of \cite{BGH-0712.3206} (in particular, Eq.~(24)), these are
$
{\cal L}_{kin} \;=\; \frac{1}{2}
\left(\frac{\delta {\cal L}}{\delta{\cal R}_{abcd}}\right)^{(0)}\nabla^{(0)}_{e}h_{ad}\nabla^{e(0)}
h_{bc} \;.
$
We have  neglected any inessential factors, as well as a term that makes no contribution
when coupled to a conserved source. The background quantities have to be evaluated on  the $\lambda=0$ solution, since
each $h$ already represents one order of $\lambda$.

The $\lambda=0$ solution is the well-known AdS $3$-brane, for which $-g_{tt}=g_{rr}= r^2\left[r^2-\frac{r_h^2}{r^2}\right]$ (with $r_h=1/(\pi T)$) and $g_{xx}=g_{yy}=g_{zz}=r^2$.
It is also useful to note that $({\cal R})^{(0)}=-20$ and $({\cal R}^{\ b}_a)^{(0)}= -4\delta^{\ b}_a$.
The calculation proceeds in a straightforward manner. Specializing to the $h_{xy}$ gravitons of interest, we obtain
$
{\cal L}_{kin} = -\frac{1}{4}\sum_{a\neq b}^{\{ x,y,z \} }
\left[\left(1-8\lambda G_E +
4\lambda G_E {\cal R}^{ab}_{\;\;\;ab}\right) h_{ab}\Box h^{ab} \right],
$
where some boundary terms have been dropped, as well as a bulk term that is not of the kinetic form,
and the usual  summation conventions should be ignored.
The leading-order term is just the standard graviton propagator prior to any gauge fixing,
and so we need  to compare the sign of the higher-order terms with this one.
To proceed, let us  evaluate the four-index Riemann tensor on the horizon, although any other radial surface would serve just as
well. In this case,
${\cal R}^{xy}_{\;\;\;xy} = 0$,  with all other relevant possibilities  being redundant. So that the extremely simple end result is (with
the summation suppressed)
$
{\cal L}_{kin} = -\frac{1}{4}\left(1-8\lambda G_E\right) h_{ab}\Box h^{ab}\;.
$
Evidently, $\lambda>0$ induces
a negative or ghost contribution to the propagator.
This agrees perfectly with the observation that $\lambda>0$ Gauss--Bonnet models
violate  the KSS bound \cite{BLMSY-0712.0805,same-day}.  The reason is clear: For $\lambda>0$,  $\eta$ picks up
a correction of $-8\lambda$  while $s$ at the horizon is found to  receive no such correction.

There is an important lesson that can be learned from the two examples. Had we first integrated out the KK modes and then { truncated} the resulting theory, the gravitational action would end up looking very similar to the
just-discussed Gauss--Bonnet model. In string theory, which is ghost free, when one  integrates out the massive modes and truncates the ensuing expansion, the leading-order result is indeed a Gauss--Bonnet theory. It may well be that, for the calculation of scattering amplitudes and some other physical quantities, such a truncation is perfectly fine. However, since the ratio $\eta/s$ is
especially sensitive to the presence of ghosts, extra care must now be taken.

{\bf Acknowledgments:}
The research of RB was supported by The Israel Science Foundation grant no 470/06.
The research of AJMM is supported by the University of Seoul.
AJMM thanks  Ben-Gurion University for their hospitality during his visit.


\begin{thebibliography}{99}

\bibitem{KSS-hep-th/0405231}       
P.~Kovtun, D.~T.~Son and A.~O.~Starinets,
``Viscosity in strongly interacting quantum field theories from black hole physics,''
Phys.\ Rev.\ Lett.\ {\bf 94}, 111601 (2005)  [arXiv:hep-th/0405231].


\bibitem{Schafer:2009dj}
  T.~Schafer and D.~Teaney,
  ``Nearly Perfect Fluidity: From Cold Atomic Gases to Hot Quark Gluon
  Plasmas,''
  arXiv:0904.3107 [hep-ph].

\bibitem{coldFG}
J.~E.~Thomas, ``Is an Ultra-Cold Strongly Interacting Fermi Gas a Perfect Fluid?,"
arXiv:0907.0140v1 [cond-mat.quant-gas]

\bibitem{Collaboration:2009yd}
  CERES Collaboration,
  ``Viscosity of the matter created in nucleus-nucleus collisions at the SPS
  measured via two-pion interferometry,''
  arXiv:0907.2799 [nucl-ex].

\bibitem{maldacena1}
  J.~M.~Maldacena,
  ``The large N limit of superconformal field theories and supergravity,''
  Adv.\ Theor.\ Math.\ Phys.\  {\bf 2}, 231 (1998)
  [Int.\ J.\ Theor.\ Phys.\  {\bf 38}, 1113 (1999)]
  [arXiv:hep-th/9711200].

\bibitem{maldacena2}
  O.~Aharony, S.~S.~Gubser, J.~M.~Maldacena, H.~Ooguri and Y.~Oz,
  ``Large N field theories, string theory and gravity,''
  Phys.\ Rept.\  {\bf 323}, 183 (2000)
  [arXiv:hep-th/9905111].




\bibitem{KSS-hep-th/0309213}      P.~Kovtun, D.~T.~Son and A.~O.~Starinets,
``Holography and hydrodynamics: Diffusion on stretched horizons,''
JHEP {\bf 0310}, 064 (2003) [arXiv:hep-th/0309213].

\bibitem{PSS-hep-th/0104066}       G.~Policastro, D.~T.~Son and A.~O.~Starinets,
``The shear viscosity of strongly coupled N = 4 supersymmetric Yang-Mills plasma,''
Phys.\ Rev.\ Lett.\ {\bf 87}, 081601 (2001) [arXiv:hep-th/0104066].


\bibitem{SS-0704.0240}                  D.~T.~Son and A.~O.~Starinets,
``Viscosity, Black Holes, and Quantum Field Theory,''
Ann.\ Rev.\ Nucl.\ Part.\ Sci.\ {\bf 57}, 95 (2007) [arXiv:0704.0240 [hep-th]]



\bibitem{BLMSY-0712.0805}      M.~Brigante, H.~Liu, R.~C.~Myers, S.~Shenker and S.~Yaida,
``Viscosity Bound Violation in Higher Derivative Gravity,''
 Phys.\ Rev.\ D {\bf 77}, 126006 (2008) [arXiv:0712.0805 [hep-th]].

\bibitem{same-day}         Y.~Kats and P.~Petrov,
``Effect of curvature squared corrections in AdS on the viscosity of the dual gauge theory,''
JHEP {\bf 0901}, 044 (2009) [arXiv:0712.0743 [hep-th]].


\bibitem{BM-0808.3498}
R.~Brustein and A.~J.~M.~Medved,
``The ratio of shear viscosity to entropy density in generalized theories of gravity,''
Phys.\ Rev.\ D {\bf 79}, R021901 (2009)
[arXiv:0808.3498 [hep-th]].


\bibitem{BGH-0712.3206}    
  R.~Brustein, D.~Gorbonos and M.~Hadad,
  ``Wald's entropy is equal to a quarter of the horizon area in units of the
  effective gravitational coupling,''
  Phys.\ Rev.\ D {\bf 79}, 044025 (2009)
  [arXiv:0712.3206 [hep-th]].




\bibitem{PSS2}
  G.~Policastro, D.~T.~Son and A.~O.~Starinets,
  ``From AdS/CFT correspondence to hydrodynamics,''
  JHEP {\bf 0209}, 043 (2002)
  [arXiv:hep-th/0205052].

\bibitem{Kovtun:2005ev}
  P.~K.~Kovtun and A.~O.~Starinets,
  ``Quasinormal modes and holography,''
  Phys.\ Rev.\  D {\bf 72}, 086009 (2005)
  [arXiv:hep-th/0506184].

\bibitem{Iqbal:2008by}
  N.~Iqbal and H.~Liu,
  ``Universality of the hydrodynamic limit in AdS/CFT and the membrane
  paradigm,''
  Phys.\ Rev.\  D {\bf 79}, 025023 (2009)
  [arXiv:0809.3808 [hep-th]].


\bibitem{BM-shear}
R.~Brustein and A.~J.~M.~Medved,
``The shear diffusion coefficient for generalized theories of gravity,''
Phys.\ Lett.\ B {\bf 671}, 119 (2009)
[arXiv:0810.2193 [hep-th]].


\bibitem{Zakharov}  V.~I.~Zakharov,
``Linearized gravitation theory and the graviton mass,''
JETP Lett.\ {\bf 12}, 312 (1970)
[Pisma Zh.\ Eksp.\ Teor.\ Fiz.\ {\bf 12}, 447 (1970)].

\bibitem{veltman}
H.~van Dam and M.~J.~G.~Veltman,
``Massive And Massless Yang-Mills And Gravitational Fields,''
Nucl.\ Phys.\ B {\bf 22}, 397 (1970).


\bibitem{Dvali1}
G.~Dvali,
``Predictive Power of Strong Coupling in Theories with Large Distance
Modified Gravity,''   New J.\ Phys.\ {\bf 8}, 326 (2006)
[arXiv:hep-th/0610013].

\bibitem{Dvali2}   G.~Dvali, O.~Pujolas and M.~Redi,
``Non Pauli-Fierz Massive Gravitons,''
Phys.\ Rev.\ Lett.\ {\bf 101}, 171303 (2008) [arXiv:0806.3762 [hep-th]].

\bibitem{wald1}
  R.~M.~Wald,
  ``Black hole entropy is the Noether charge,''
  Phys.\ Rev.\  D {\bf 48}, 3427 (1993)
  [arXiv:gr-qc/9307038].


\bibitem{wald2}
 V.~Iyer and R.~M.~Wald,
  ``Some properties of Noether charge and a proposal for dynamical black hole
  entropy,''
  Phys.\ Rev.\  D {\bf 50}, 846 (1994)
  [arXiv:gr-qc/9403028].


\bibitem{wald3}
  T.~Jacobson, G.~Kang and R.~C.~Myers,
  ``On Black Hole Entropy,''
  Phys.\ Rev.\  D {\bf 49}, 6587 (1994)
  [arXiv:gr-qc/9312023].



\end{thebibliography}
\end{document}